\documentclass[review]{elsarticle}

\usepackage{hyperref}
\usepackage{amsmath}
\DeclareMathOperator*{\argmin}{arg\,min}

\usepackage{xcolor}
\usepackage{tcolorbox}
\usepackage{caption}
\definecolor{mydiffcolor}{rgb}{0,0,0}

\journal{Applied Energy}

\begin{document}

\begin{frontmatter}

\title{A physics-informed data reconciliation framework for real-time electricity and emissions tracking}
\tnotetext[mytitlenote]{A short version of this paper was presented at Applied Energy Symposium: MIT A+ B, August 13-14, Boston. This paper is a substantial extension of the short version of the conference paper.}


\author[stanfordaddr]{Jacques A. de Chalendar\corref{mycorrespondingauthor}}

\cortext[mycorrespondingauthor]{Corresponding author}
\ead{jdechalendar@stanford.edu}

\address[stanfordaddr]{Department of Energy Resources Engineering, Stanford University, Stanford, CA 94305}

\author[stanfordaddr]{Sally M. Benson}

\begin{abstract}
\textcolor{mydiffcolor}{
To encourage and guide decarbonization efforts, better tools are needed to monitor real-time CO$_2$ and criteria air pollutant emissions from electricity consumption, production, imports, and exports. Using real-time data from the electricity system is especially challenging for quantitative applications  requiring high quality and physically consistent data. Until now,  time-intensive, ad-hoc and manual data verification and cleaning strategies have been used to prepare the data for quantitative analysis. As an alternative to existing techniques, here we provide a physics-informed framework to greatly accelerate and automate data processing to enable internally consistent electric system consumption, production, import, and export data in near real-time. A key component of this framework is an optimization program to minimize the data adjustments required to satisfy energy conservation equations. The effectiveness of this method is demonstrated by applying it to the continental United States electricity network. The resulting publicly-available data set, which provides in near-real time, hourly updates on electricity generation, consumption, imports, exports and associated emissions, is the first of this nature.}
\end{abstract}


\end{frontmatter}


\section{Introduction}
\textcolor{mydiffcolor}{As efforts to curb greenhouse gas emissions intensify, tracking emissions from energy consumption, production, imports, and exports in a timely manner will be critical for reducing emissions. New tools that are efficient, robust, and reliable are needed to measure and analyze emissions at increased spatial and temporal resolution.}
\par The electric sector is a prime target for these efforts because of its large and growing share of energy use, associated emission of greenhouse gases (GHGs), \textcolor{mydiffcolor}{and critical role in decarbonizing the heating and transportation sectors.}  Today, emissions from electricity and heat generation contribute about 41\% of the world's 32.8 Gtons of carbon dioxide (CO$_2$) emissions from fuel combustion \cite{iea2019co2}. In the United States (US), electricity accounts for about 28\% of GHG emissions \cite{EPA2018}. In the future, electricity consumption is expected to increase rapidly with electrification of transportation, heating, and industry \cite{davis2018net}. Reliable and high quality electricity system data will be critical for understanding current operations and tracking changes in a rapidly evolving grid. The availability of high quality, near real-time data also provides policy makers with new tools to reduce GHG emissions through carbon-aware pricing or other incentives for using electricity when the emission intensity is low. 

\par \textcolor{mydiffcolor}{Unlike the past when electricity was largely provided by thermal generating units and hydropower resources, the carbon intensity of electricity now varies widely in space and time. In many regions of the United States, large shares of wind and solar power electricity dominate the grid during portions of the day. During these times, the carbon intensity of the grid drops to very low values. This is expected to increase at an accelerated pace due to efforts to decarbonize the electricity sector. To compensate for variable generation from wind and solar power, other resources are used to meet electricity demand,   including the use of flexible generation from gas and coal plants, large scale energy storage, and electricity exchanges between balancing areas. Together these new grid operating and balancing strategies result in the carbon intensity of the grid now changing dynamically on time scales spanning hours, to day, seasons, and years.} Under these conditions, tracking the carbon intensity of the grid is more challenging than ever. In a world demanding more accurate and timely GHG accounting, new tools are needed to provide this information when and where it is needed.

\par Timely availability of data on grid carbon intensity will enable accelerated decarbonization of a number of sectors. Real-time pricing of electricity modulated by grid carbon intensity would encourage use of the electricity when and where the grid is the cleanest \cite{allcott2011rethinking}. Benefits of transportation electrification can be quantified and optimized to account for local variations in carbon intensity \cite{tamayao2015regional}. The value of credits in carbon markets for the industrial and commercial sector can be rigorously documented through real-time accounting for time-of-use electricity consumption and emissions \cite{dechalendar2019100}.

\par Electricity system planning studies can also benefit from high quality operating and carbon intensity data. Options for future electricity systems with increased shares of renewable generation, such as large-scale storage or increased transmission, are often assessed using capacity expansion models; see \textit{e.g.} \cite{de2016value} or \cite{macdonald2016future}. Capacity expansion studies typically rely on optimization programs to search for the lowest cost investment and operating strategies for electrical power systems, as well as on a wide range of data inputs, from historical information on electric system operations to the costs of different technologies \cite{loulou2004documentation, eia2014electricity, short2011regional, fripp2012switch}. Accurate information about power and associated emission exchanges are needed to both calibrate and parameterize such models.
\par \textcolor{mydiffcolor}{Computation of emissions embodied in electricity systems is an active field and a key motivation for the framework introduced in this paper. A growing body of work is concerned with understanding how emissions and carbon intensity vary by location, season or time of day \cite{siler2013regional, zivin2014spatial}. The focus of the field is not limited to the US, recent work includes examples that focus on the European \cite{tranberg2019real, schafer2020tracing} and Chinese \cite{qu2017co2} contexts.}

\par Whether it is to provide near-real time information to incentivize GHG emission reductions, provide accurate emissions accounting from an electrified transportation sector, to feed capacity expansion modeling efforts, or for other electric sector-related studies, procuring reliable data of sufficient quality is critical to the subsequent analyses. For example, the methodology previously developed by the authors to compute hourly production- and consumption-based emissions in the US electricity system relied on a fully coupled economic multi-regional input-output (MRIO) model \cite{de2019tracking}. Computing emissions through this approach requires solving a linear system that can become ill-conditioned when the input data are not physically consistent (\textit{e.g.} if energy conservation is violated).

\par This work introduces a framework to automatically reconcile inconsistent data on electricity generated, consumed and exchanged by a set of regions connected by an electricity network. The goal of the framework is two-fold: (i) to provide an automated assessment of data quality and internal consistency and (ii) to provide an educated guess for a corrected data set. A central component of this framework is an optimization program that models the internal consistency relations that are to be expected between different variables in the data set and minimizes the data adjustments needed to satisfy those relations.

\par The performance of the method is analyzed using a historical data set. Two illustrative examples are also provided to demonstrate how the new framework enables researchers to now perform similar analyses in near real-time and to continuously monitor emissions rates in the US electricity system. The raw and cleaned US electricity data as well as the consumption- and production-based emissions that are generated as part of the illustration are streamed to a publicly available data collection service \cite{de2021data}. The framework that is developed is not specific to the US, however, and can be readily applied to electricity grids in other locations.
\par Although the physics-informed data reconciliation methodology was developed with an electricity and emissions tracking application in mind, we believe it will benefit other consumers of electric system operating data as well, \textit{e.g.} researchers attempting to create realistic simulations of electricity systems in the context of capacity expansion modeling, as well as policy makers and private sector actors.
\par \textcolor{mydiffcolor}{To the best of our knowledge, data cleaning strategies for similar energy system data are most often manual, ad-hoc and time-consuming. This work is novel and unique in that it introduces a systematic framework to accelerate and automate such tasks, which will be valuable to policy makers and private sector actors, as well as enable further research in the field.}

\section{Data and methods}
\label{sec:2}
In the data reconciliation framework introduced in this work, the electricity system is modeled as a graph whose nodes are regions of the electric grid and whose edges are transmission lines between regions. \textcolor{mydiffcolor}{A key component of this framework is an optimization program that minimizes the adjustments to input data such that energy conservation equations are satisfied.} The relevant electric system operating data correspond to information on electricity produced and consumed in each node and exchanged along each edge of this graph. Optionally, the data reconciliation framework can also flexibly accommodate data on electricity production by source. Hourly data such as these are increasingly made publicly available, \textit{e.g.} for the US electricity system by the US Energy Information Administration (US EIA) \cite{eia2018bulk}, or for the European electricity system by the European Network of Transmission System Operators (ENTSO-E) \cite{entsoe2020platform}. The methodology was developed for and applied to data from the US EIA, but can be readily adapted to data from other locations.

\textcolor{mydiffcolor}{
\paragraph{Data pre-processing}
In the data pre-processing step, missing data fields are added, unrealistic values are rejected, and a first guess is provided for missing and rejected values. In the application that is the main focus of this work, the heuristics that are described next have been found to result in good quality reconciled data sets. We note that in other applications, or in cases where the user has more specific information, \textit{e.g.} about the quality of one of the data fields, these heuristics can be replaced with user-defined heuristics.}
\par Application of the framework requires that information be provided for each node of the associated graph. The first step therefore consists in adding missing data fields. In the US, for example, some regions only generate electricity and therefore report no demand. Canada and Mexico only report interchange data, and some trade links are only reported by one of the two trading partners. For each region, values for the missing fields are inferred from other fields and physical equations (\textit{e.g.} conservation of energy). Non-US regions are treated as if they were generation-only regions generating enough to meet their exports.
\par Heuristic filters are then applied to reject data values that appear unrealistic. This is done in two ways: (i) using static thresholds and (ii) using a dynamic threshold that is based on computing a ten-day moving average and standard deviation. Values that are farther than four standard deviations from the mean are rejected (if the data were normal, 99.9\% of the data should fall within four standard deviations of the mean). We note that other filtering methods could also be applied at this step, whose goal is to reject bad data. This is also the step where a user with more targeted information on the quality of different data fields can make manual data adjustments or provide a custom filtering heuristic.
\par \textcolor{mydiffcolor}{A second set of heuristics is used to replace missing and rejected data values with a reasonable first guess. In our implementation, we use linear interpolation for each hour of the day between the nearest days with valid data where possible, and propagate the last valid data values forwards or backwards otherwise. Said in another way, we use linear interpolation between the last valid data point for the hour of interest and the next valid data point for that hour.}

\paragraph{Optimization-based data reconciliation}
The main component of the methodology is to compute the minimal data adjustments needed to ensure that the post-processed data set is internally consistent, as expressed by a set of physical relations that guarantee energy conservation and other basic properties hold (such as non-negativity of physical variables). This procedure is performed independently for each time step. In the following, we call $D_r$ demand, $G_r$ generation, $T_r$ total interchange for region $r$, where $\mathcal{R}$ is the set of regions; $\tau_{r_1,r_2}$ the exchange from $r_1$ to $r_2$; $P_{r,s}$ the generation in region $r$ by source $s$, where $\mathcal{S}_r$ is the set of generation sources in region $r$. These are all data inputs and are treated as \textit{parameters} in the optimization program. The \textit{decision variables} of the optimization program are the adjustments made to each of the data inputs, called  $\delta_{X,i}$ where $X$ is one of the data inputs and $i$ is used to index in the set for that input. Table~\ref{tab:nomenclature} can be used as a reference for notation.
\par With this notation, the data reconciliation optimization program can be written as:
\begin{subequations}
\begin{align}
\Delta=\argmin \quad & \sum_{r\in\mathcal{R}}{w_{D,r}\delta_{D,r}^{2} +w_{G,r} \delta_{G,r}^2+w_{T,r}\delta_{T,r}^2}\nonumber\\
&+ \sum_{(r_1,r_2)\in \mathcal{R}^2}{w_{\tau,r_1, r_2}\delta_{\tau,r_1,r_2}^2} + \sum_{r\in \mathcal{R}}{\sum_{s\in\mathcal{S}_r}w_{P,r,s}\delta_{P,r,s}^2},\\
\textrm{s.t.} \quad & D_{r}+ \delta_{D,r} + T_{r} +\delta_{T,r} -G_r - \delta_{G,r} = 0, \quad r\in\mathcal{R}\label{eq:emissions_cleaning:e},\\
& G_r + \delta_{G,r} - \sum_{s\in\mathcal{S}_r} (P_{r,s}+ \delta_{P,r,s})= 0, \quad r\in\mathcal{R}\label{eq:emissions_cleaning:b},\\
& T_{r_1} + \delta_{\tau ,r_1} - \sum_{r_2\in\mathcal{R}} (\tau_{r_1,r_2}+ \delta_{\tau,r_1,r_2})= 0, \quad r_1\in\mathcal{R}\label{eq:emissions_cleaning:c},\\
& \tau_{r_2,r_1}+ \delta_{\tau,r_2,r_1} + \tau_{r_1,r_2}+ \delta_{\tau,r_1,r_2}= 0, \quad (r_1,r_2)\in\mathcal{R}^2\label{eq:emissions_cleaning:d},\\
& D_{r}+ \delta_{D,r}\geq \epsilon, \quad r\in\mathcal{R}\label{eq:emissions_cleaning:f},\\
& G_{r}+ \delta_{G,r}\geq \epsilon, \quad r\in\mathcal{R} \label{eq:emissions_cleaning:g},\\
& P_{r,s}+ \delta_{P,r,s}\geq \epsilon, \quad s\in\mathcal{S}_r, r\in\mathcal{R} \label{eq:emissions_cleaning:h}.
\end{align}
\label{eq:emissions_cleaning}
\end{subequations}
\begin{table}[tbhp]
    \centering
    \begin{tabular}{ll}
    \multicolumn{2}{l}{\textbf{Sets}} \\\hline
        $\mathcal{R}$ & Set of balancing authorities (regions) in the US electricity grid. \\
        $\mathcal{S}_r$ & Set of generation sources in region $r$. \\\\
    \multicolumn{2}{l}{\textbf{Parameters}} \\\hline
    $D_r$ & Electricity consumed in region $r$ (MWh)\\
    $G_r$ & Total electricity generated in region $r$ (MWh)\\
    $P_{r,s}$ & Electricity generated from source $s$ in region $r$ (MWh)\\
    $T_{r}$ & Total electricity interchange for region $r$ (MWh). By convention, exports\\
    & are positive and imports negative.\\
    $\tau_{r_1,r_2}$ & Electricity transferred from region $r_1$ to region $r_2$ (MWh).\\
    $w_{X,i}$ & Objective function weight for parameter $X$, indexed by $i$, where \\
    & $X\in\{D,G,P,T,\tau\}$ (dimensionless).\\
    $\epsilon, A, \gamma$ & Numerical constants (1 MWh, 10 GWh, 100 MWh).\\\\
    \multicolumn{2}{l}{\textbf{Variables}} \\\hline
    $\delta_{X,i}$ & Adjustment variable for parameter $X$ indexed by $i$ (MWh).\\
    $\Delta$ & Vector formed by concatenating all of the adjustment variables.\\
    \end{tabular}
    \caption{Nomenclature for the data reconciliation optimization program~\ref{eq:emissions_cleaning}.}
    \label{tab:nomenclature}
\end{table}

\par The optimization program~\ref{eq:emissions_cleaning} minimizes the weighted Euclidean norm of the adjustments subject to constraints that model key constitutive physical relations. In particular, the adjustments $\Delta$ computed by solving this optimization program guarantee that the adjusted data set will be physically consistent for each region, \textit{i.e.} that the sum of demand and total interchange matches generation for each node~\ref{eq:emissions_cleaning:e}; that the sum of generation from each source matches total generation~\ref{eq:emissions_cleaning:b}; that total interchange for a node matches the sum of what is exchanged with other regions~\ref{eq:emissions_cleaning:c}; that the exchange matrix is anti-symmetric~\ref{eq:emissions_cleaning:d}; that energy is conserved~\ref{eq:emissions_cleaning:e}; and that electricity consumed and produced are positive~\ref{eq:emissions_cleaning:f} --~\ref{eq:emissions_cleaning:h}. $\epsilon$ is a very small constant that controls the algorithm's precision and is chosen to be 1 MWh in the numerical implementation. In the case where data on generation by source is not available, the corresponding variables and constraints are simply omitted (in the data set used for illustration, this is the case for data before July 1\textsuperscript{st}, 2018).

\par The weights $w_{X,i}$ play a key role in ranking the different data fields according to where larger adjustments are acceptable. In practice, penalties should be stronger for parameters that have smaller absolute values. In the numerical illustration for this paper, the heuristic
\begin{equation}
    w_{X,i} = \frac{A}{\max (|\Xi_i|, \gamma)},
    \label{eq:cleaning_weights}
\end{equation}
is used, where $\Xi_i$ is the ten-day rolling average for $X_i$, computed when rejecting values in the data pre-processing step, and the constants $A, \gamma$ are chosen such that weights remain in an acceptable range (with numerical values of 10 GWh and 100 MWh, respectively, weights are observed to remain between 1 and 1,000). The expression $\max (|\Xi_i|, \gamma)$ corresponds to the absolute value of the rolling mean for $X_i$, truncated at $\gamma$ to ensure that the denominator of equation~\ref{eq:cleaning_weights} remains large enough. \textcolor{mydiffcolor}{The adaptive weights chosen by this heuristic are inversely proportional to this typical value, so that data fields with large values will tolerate larger adjustments than fields with small values. If more specific information on the data quality of certain data fields is available, the user can specify custom weights to the optimization program. These weighing factors make adjustments to those data fields very expensive and steer the solver towards a solution that leaves these data fields largely untouched.} In combination with the manual input of corrected data, these weights can be tuned, making the data reconciliation framework very flexible.

\par We note that the size of the decision vector for the optimization program scales with at most the square of the number of nodes in the data set, since there are at most $|\mathcal{R}|(|\mathcal{R}|-1)$ trade adjustments $\delta_{T,r}$. In the example application that is used to analyze the performance of the method in section 3, there are 68 regions. The optimization program can be solved in seconds on a laptop computer with off-the-shelf quadratic solvers, so computational cost is not currently a concern.

\par The data reconciliation framework developed in this work is applied to publicly available data downloaded from the US EIA online data facility from 2015 onwards (Form EIA-930) \cite{eia2018bulk} to generate a historical data set. This dataset provides electric system operating data on generation, consumption and exchanges of electricity for every hour and at the level of the balancing area (BA). In the remainder of this paper, BAs will be referred to as ``regions'' to simplify language. A full table of abbreviations for the different regions in the US and a reference map to situate regions can both be found online \cite{eia2021acronyms, eia2021map}. Since July of 2018, the EIA online data facility also releases the hourly generation mix in each region. Data transformed through our framework is released to a public location. New data is queried from the US EIA Application Programming Interface (API) on an hourly basis, transformed, and uploaded to the same public location. A small website is maintained to access the data and track the most recent data that are available \cite{de2021data}.

\par Two implementations of the optimization program (Eq. ~\ref{eq:emissions_cleaning}) are provided with the supplemental code \cite{de2021code} for this paper: a first version using the Pyomo modeling library \cite{hart2017pyomo} and solved with Gurobi \cite{GurobiOptimizationIncorporated2018}, and a second (slower) version using only open-source tools \cite{diamond2016cvxpy}. The second implementation is used to maintain the data set that is publicly released \cite{de2021data}.

\par \textcolor{mydiffcolor}{Directly using the US EIA electric system operating data can be challenging due to numerous inconsistencies in the data \cite{eia2019issues}. For example, the interchange reported by region $r_1$ to or from region $r_2$ does not match the corresponding interchanges reported for region $r_2$. These inconsistencies are particularly problematic when computing consumption-based emissions because the MRIO framework relies on solving linear systems.} When the data supplied to our solver is inconsistent, the linear system is often ill-conditioned and solving it is prone to numerical instability.

\paragraph{Production- and consumption-based emissions}
For completeness, we briefly outline our method for computing production- and consumption-based emissions here and refer to \cite{de2019tracking} (and references therein) for a more in-depth description of consumption-based accounting for electricity system operating data. In this framework, pollution is embodied in generated electricity and subsequently flows through the electricity network. Produced emissions are identified by the administrative territory (balancing area) in which they are physically emitted. Consumption-based emissions are defined by the administrative territory in which electricity is consumed, and we refer to them as ``consumed'' emissions. We similarly refer to ``traded'' emissions as the emissions embodied in hourly electricity exchanges.
\par We call $f_s$ the emissions factor for generation source $s$ and compute the emissions produced in region $r$ as:
\begin{equation}
    F_r = \sum_{s\in\mathcal{S}_r} f_s P_{r,s}, \quad r\in\mathcal{R}
\end{equation}
For the results presented in this paper, we use the most recently available life-cycle analysis estimates from the IPCC~\cite{IPCC2011} as the technology-specific emissions factors (see Supplementary Information). The same procedure as in \cite{de2019tracking} is then followed to compute consumption-based emissions. Following the US EIA API's convention, $\tau_{r_1,r_2}$ corresponds to the electricity sent from $r_1$ to $r_2$, is negative for imports and positive for exports. Imports into $r_1$ from $r_2$ are computed as $u_{r_1,r_2}=-\min (\tau_{r_1,r_2},0)$ and the following linear system can be written to compute the consumption-based emissions factor in region $r$, $x_r$:
\begin{equation}
    x_r(G_r+U_r)-\sum_{r_2} x_{r_2} u_{r,r_2} = F_r, \quad r\in\mathcal{R}
\end{equation}
Note that this last equation corresponds to equation (4) from \cite{de2019tracking} and provides information on how embodied emissions propagate through the electric grid, from production to consumption. Also note that (10) accounts for trans-shipments of electricity across regions. $U_r$ corresponds to the total imports for region $r$, $U_r=\sum_{r_2}u_{r,r_2}$.

\section{Results}
\textcolor{mydiffcolor}{The main contribution from this work is a physics-informed data reconciliation framework that enables real-time accounting for electricity generation, consumption, imports, exports and associated carbon or air pollution emissions across interconnected balancing areas. The performance of this framework is analyzed in detail on a historical data set in section 3.1, and further tested against artificial erroneous data in section 3.2. Sections 3.3 and 3.4 provide two illustrative examples of analyses that are enabled by the framework.}

\subsection{Automated data reconciliation}
The performance of the data reconciliation framework is analyzed using a historical data set from July 1\textsuperscript{st}, 2018 to January 21\textsuperscript{st}, 2021. Summary results are presented in this section and exhaustive region by region reports are provided in the Supplementary Information for this paper.

\par
\begin{figure}[tbhp]
\centering
\includegraphics[width=\textwidth]{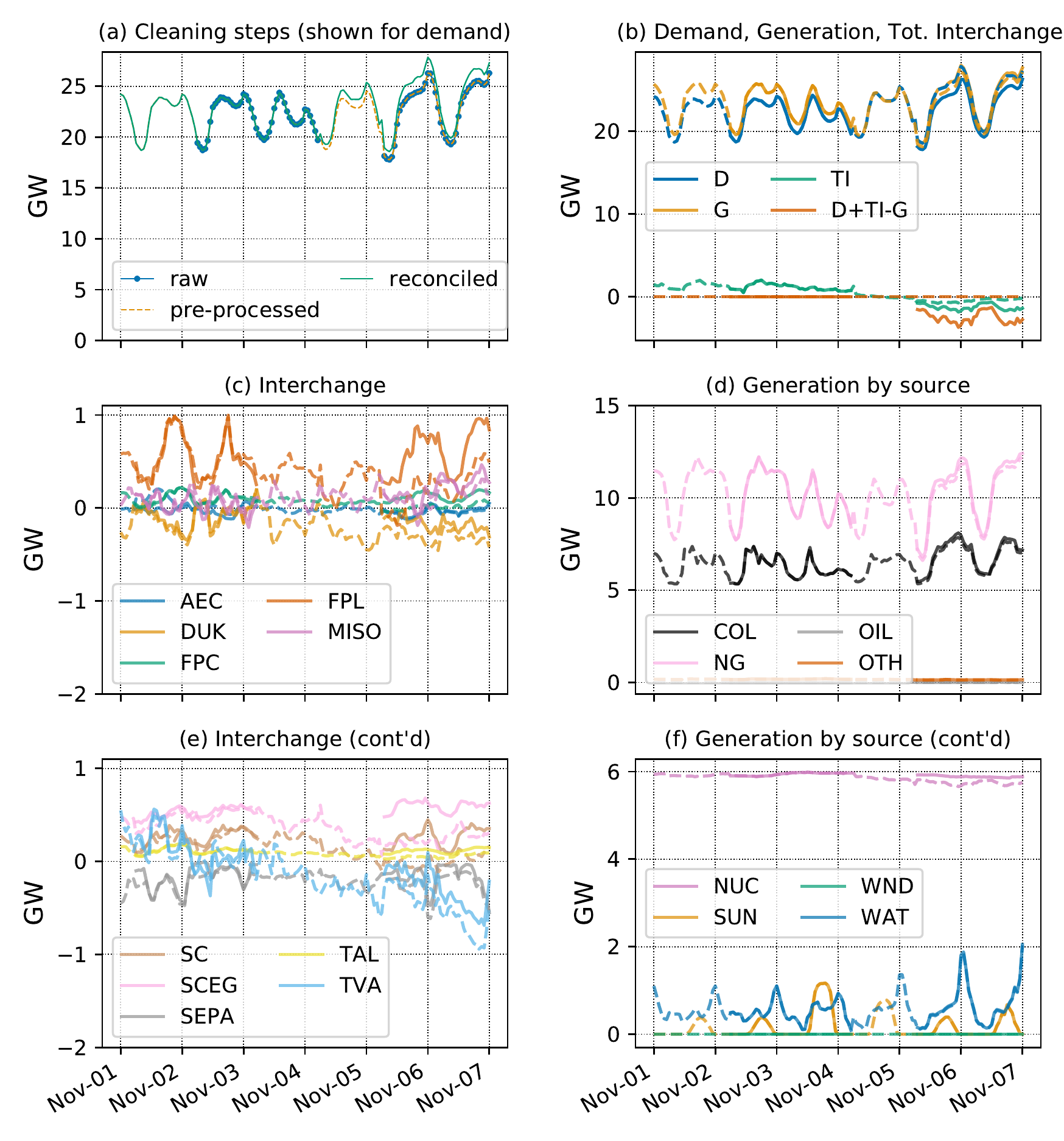}
\caption{Data reconciliation results for a week in November of 2018 for Southern Company Services, Inc. - Trans (SOCO). (a) Data for demand at each stage of the reconciliation process. (b) Demand, generation and total interchange data at the raw (full line) and reconciled (dashed line) stages. The energy balance relation is also shown and is zero for the internally consistent reconciled data. (c) and (e) Interchange data at the raw (full line) and reconciled (dashed line) stages. (d) and (f) Generation by source data at the raw (full line) and reconciled (dashed line) stages.}
\label{fig:1}
\end{figure}

 \textcolor{mydiffcolor}{The data reconciliation framework is first illustrated for the region of the U.S. electricity system managed by Southern Company (SOCO) for a week in November 2018 (\autoref{fig:1}). In this example, all of the data fields for this region are missing from November 1st to 2\textsuperscript{nd} and from November 4\textsuperscript{th} to 5\textsuperscript{th}. Using the approach described in Section 2, a first guess for the missing values is provided at the end of the data pre-processing step.} This first guess is then adjusted during the optimization-based data reconciliation step using data from other fields and other regions. The same process is applied to other data fields for the SOCO region, and the data reconciliation framework is able to recreate realistic trajectories for the missing data. The reconciled data set satisfies the physical relationships that are expected between the different data fields, as illustrated for the relation between demand, generation, and total interchange in \autoref{fig:1}b, where the full red line shows the mismatch between demand, net interchange and generation for the raw data and the dashed red line shows the same mismatch for the reconciled data. In this example, raw data was unavailable to compute the mismatch for Nov 1\textsuperscript{st} and 4\textsuperscript{th}. The energy balance was satisfied in the raw data set for the period from Nov 2\textsuperscript{nd} to 4\textsuperscript{th}, but not from Nov 5\textsuperscript{th} to 7\textsuperscript{th}. In the reconciled data set, the energy balance was satisfied for the entire time period.

\par
\begin{figure}[tbhp]
\centering
\makebox[\textwidth][c]{\includegraphics[width=1.1\textwidth]{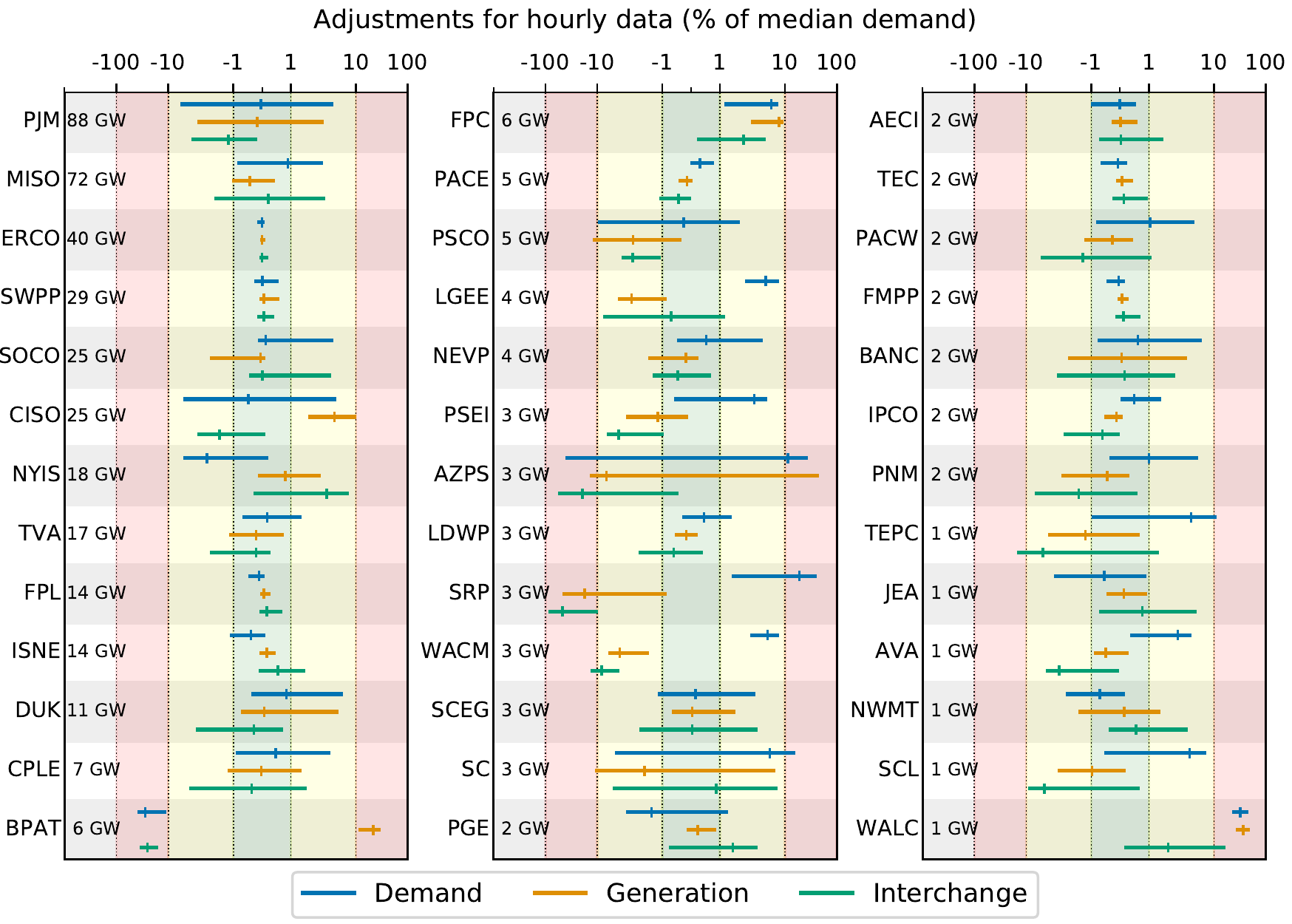}}
\caption{10\textsuperscript{th}, 50\textsuperscript{th} and 90\textsuperscript{th} percentile of adjustments to hourly data for demand, generation and total interchange, expressed as a percentage of median hourly demand for each region. Median hourly demand for each region is shown on the left of each plot next to the region name. Summary statistics are shown for the period from July 1\textsuperscript{st}, 2018 to January 21\textsuperscript{st}, 2021.}
\label{fig:2}
\end{figure}

\begin{figure}[tbhp]
\centering
\makebox[\textwidth][c]{\includegraphics[width=1\textwidth]{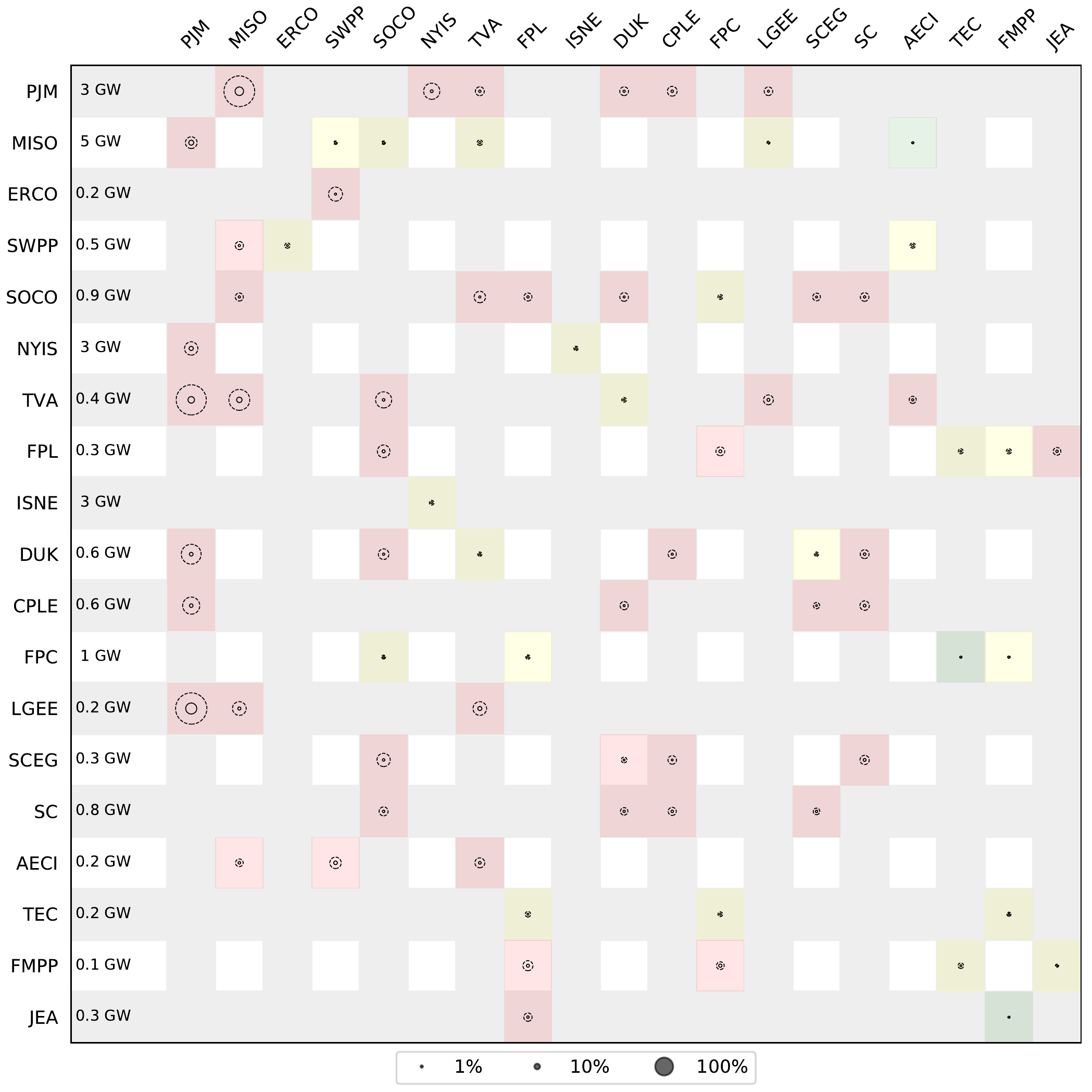}}
\caption{10\textsuperscript{th} (full circles) and 90\textsuperscript{th} (dashed circles) percentiles of the absolute value of data adjustments for inter-region electricity exchanges, expressed as a percentage of median total interchange for each region, shown to the right of the y-axis. Areas of the plot are shaded green, yellow, red for data corrections that are within $\pm$1\%, $\pm$10\%, $\pm$100\% in absolute value. Note that the data correction matrix is non-symmetric, e.g. the algorithm makes a stronger adjustment to PJM-MISO than to MISO-PJM.}
\label{fig:3b}
\end{figure}

\begin{figure}[tbhp]
\centering
\makebox[\textwidth][c]{\includegraphics[width=1.\textwidth]{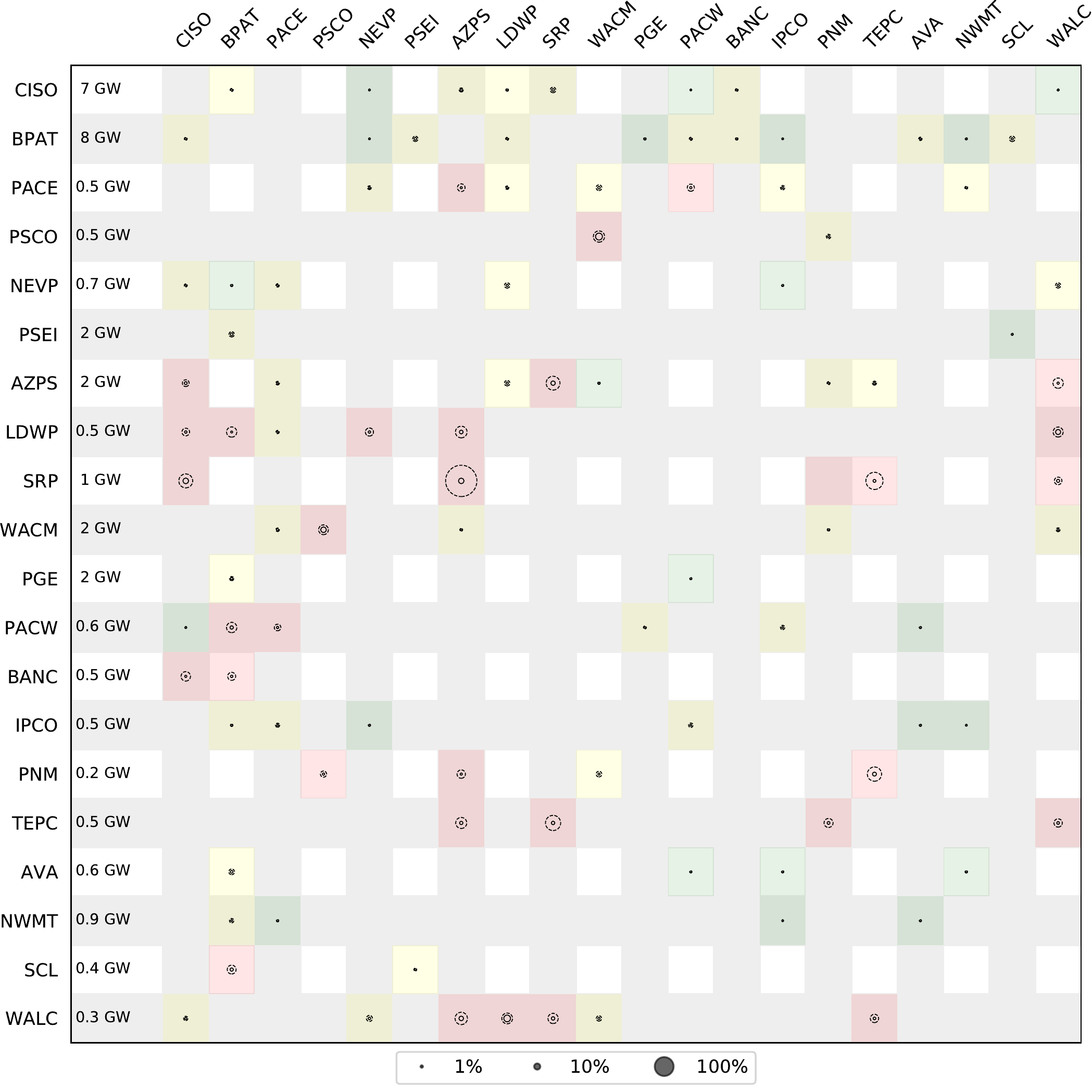}}
\caption{10\textsuperscript{th} (full circles) and 90\textsuperscript{th} (dashed circles) percentiles of the absolute value of data adjustements for inter-region electricity exchanges, expressed as a percentage of median total interchange for each region, shown to the right of the y-axis. Areas of the plot are shaded green, yellow, red for data corrections that are within $\pm$1\%, $\pm$10\%, $\pm$100\% in absolute value. Note that the data correction matrix is non-symmetric, e.g. the algorithm makes a stronger adjustment to PJM-MISO than to MISO-PJM.}
\label{fig:3a}
\end{figure}

\begin{figure}[tbhp]
\centering
\makebox[\textwidth][c]{\includegraphics[width=1.2\textwidth]{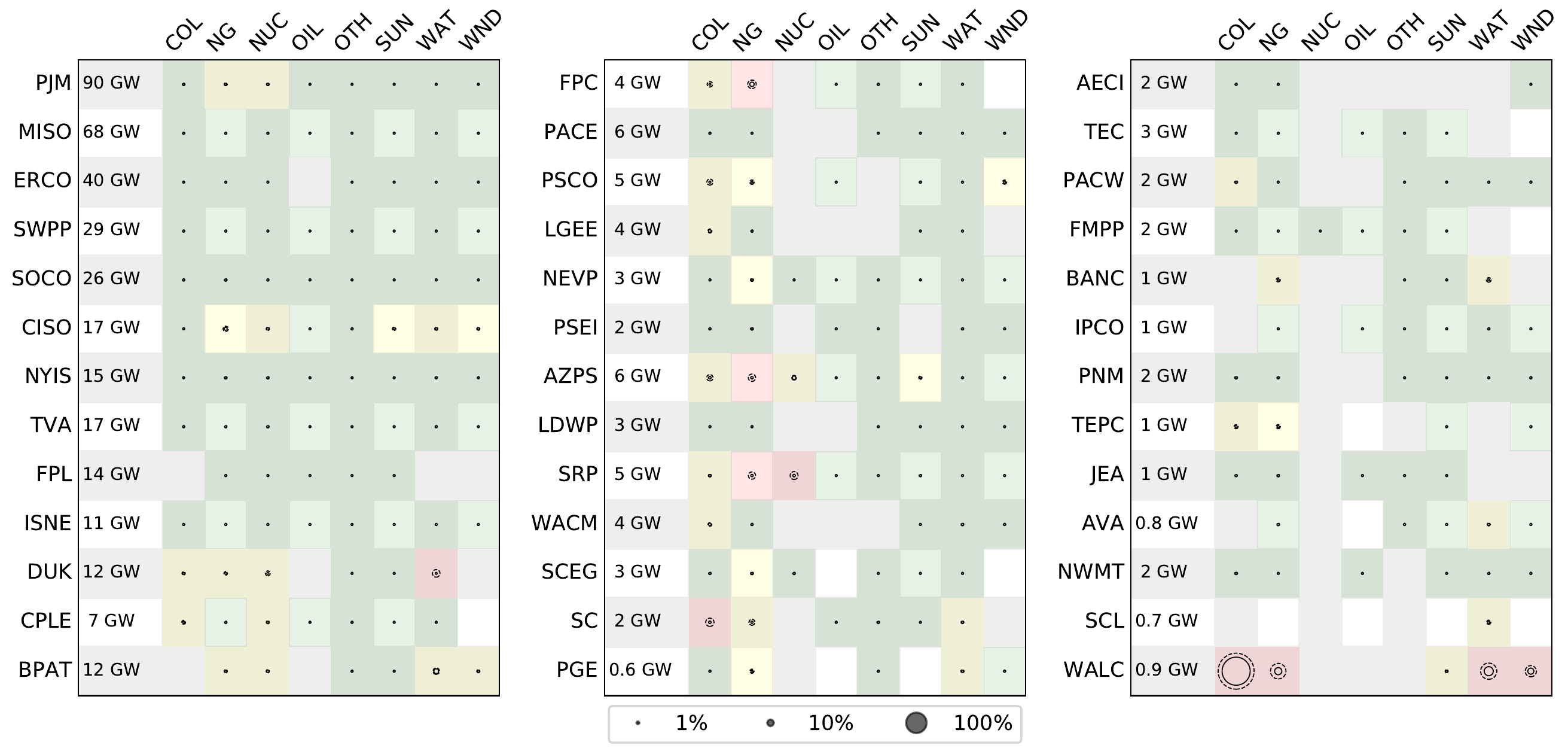}}
\caption{10\textsuperscript{th} (full circles) and 90\textsuperscript{th} (dashed circles) percentiles of the absolute value of data adjustments for the data on generation by energy source expressed as a percentage of median total generation for each region, shown to the right of the y-axis. Areas of the plot are shaded green, yellow, red for data corrections that are within $\pm$1\%, $\pm$10\%, $\pm$100\% in absolute value.}
\label{fig:4}
\end{figure}

 More comprehensive results for the hourly data adjustments that are applied by the data reconciliation framework are presented next, for the 39 regions with the largest median demand. A summary analysis of adjustments to the demand, generation and total interchange fields is presented in \autoref{fig:2}. Adjustments to exchange data can be investigated through \autoref{fig:3b} and \autoref{fig:3a}, that present results for regions in the Eastern Interconnect (EIC) and the ERCO region, and in the Western Electricity Coordinating Council (WECC), respectively. A similar analysis of adjustments to data on generation by source is presented in \autoref{fig:4}. Finally, the four supplemental reports provide an exhaustive analysis, region by region. In these supplemental reports, the medians of weekly data are shown as full lines, and the weekly 10\textsuperscript{th} to 90\textsuperscript{th} percentiles of the data are shown as the colored area around the full lines. Reports are provided for the raw and reconciled data as well as for the adaptive weights used by the data reconciliation framework and the resulting data adjustments. 

\par For some regions like the Southwest Power Pool (SWPP), Electric Reliability Council of Texas (ERCO), Florida Power and Light (FPL) or Tennessee Valley Authority (TVA), data adjustments that satisfy physical consistency are very small, and almost always within $\pm$1\% of the median demand for that region (area shaded in green). The detailed reports in the Supplemental Information show that for some of these, \textit{e.g.} ERCO, the adjustments are indeed always very small (all adjustments fall in the $[-400, +300]$ MWh range for ERCO), while for others, there are infrequent deviations that correspond to anomalies in reporting, \textit{e.g.} for SWPP. The inconsistencies caused by these infrequent deviations are resolved by the reconciliation framework.

\par For a second category of regions, data adjustments remain within $\pm$10\% of the median demand for that area (area shaded in yellow). In the largest US region, for example, the Pennsylvania, Jersey, Maryland Power Pool (PJM), adjustments can be quite large, up to almost 10 GWh for demand and generation. The behavior of the reconciliation algorithm for PJM data in the first quarter of 2020 is a good example of how the optimal adjustments that are reached can be used to diagnose data inconsistencies. In the following, we once again refer to the figures in the supplementary material \textcolor{mydiffcolor}{(figures for PJM are provided on the first page of each of the supplemental reports)}. For the first few weeks of 2020, PJM data on generation by source are much lower than the reported generation and demand. The reconciliation algorithm chooses to decrease both demand and generation by close to 10 GWh during that period, while increasing the generation produced by each source (in particular the values for coal, nuclear and natural gas are consistently increased by 3 to 6 GWh during those weeks). In the second part of the quarter, a different pattern is observed. The algorithm assigns negative adjustments to demand, but positive adjustments to generation. The adjustments for generation by source and total interchange remain small, suggesting that the issue is now a mismatch between demand and generation, which is confirmed by the supplemental report on raw data. In both cases, the data reconciliation algorithm is able to navigate these different issues and suggests plausible adjustments to reconstruct the electric system operating data.

\par The California Independent System Operator (CISO) presents another example in \autoref{fig:2} of a region where adjustments are significant but remain within $\pm$10\% of median demand (yellow area). The raw data report \textcolor{mydiffcolor}{(page 6)} shows that the violation in the energy conservation was consistently large for a contiguous period from October 2019 to August 2020, resulting in an adjustment for generation whose weekly median remained at or above 2 GWh for most of that period. Overall, adjustments to demand varied more within a week and from week to week than adjustments to generation. Adjustments to demand and total interchange were typically negative, while adjustments to generation were typically positive, consistent with the fact that CISO is a net importer of electricity.

\par A third category of regions are those for whom the adjustments in \autoref{fig:2} are quite large, sometimes farther than $\pm$100\% of the median demand for that region. Examples of regions such as these are the Bonneville Power Administration (BPAT),  Arizona Public Service Company (AZPS) or the Salt River Project (SRP). In the case of BPAT, a need for large adjustments can be seen immediately from the raw data report \textcolor{mydiffcolor}{(page 13)}: there is a consistent mismatch in reported data for demand, generation, and total interchange. The energy conservation relation is violated by up to 5 GWh for large portions of the study period.

\par AZPS and SRP are neighboring regions. While SRP reports significant exchanges with AZPS (a weekly median of 3-4 GWh), the corresponding number reported by AZPS is much lower (0-1 GWh). Since June of 2020, however, AZPS now reports a number that is much closer to the SRP number. The detailed report for the adjustments shows that without more information, the data reconciliation algorithm searches for a compromise to resolve the discrepancy: both the AZPS-SRP and SRP-AZPS exchange fields are modified so that they match. We note that more accurate information can be incorporated here: if it becomes known for a fact, for instance, that the data reported by SRP is correct and the data reported by AZPS is not, the weight of the corresponding data fields can be modified to steer the optimization program's solution towards larger adjustments for the data reported by AZPS. The AZPS example also shows that multiple sources of discrepancy can co-exist, since the energy balance constraint is also strongly violated by the raw data between November 2019 and June 2020. Once again, the reconciliation algorithm handles the different sources of discrepancy simultaneously.

\par As the share of variable renewable generation grows, there is reason to believe that exchanges will play an increasing role in the US electricity network, and procuring timely, reliable data on exchanges will become critical. Adjustments to the exchanges can be very non-symmetric, as can be observed for example in \autoref{fig:3b} for the link between PJM and MISO, the two largest regions in the US, whose median hourly exchanges represent between 2.8 and 4.0 GWh according to the raw data. Our reconciled value of 2.9 GWh appears to indicate that reporting from MISO is more accurate than reporting from PJM. Adjustments for the PJM exchange data are quite large with each of their trading partners.

\par In the current U.S. electricity system, exchanges represent a large fraction of demand in the WECC, particularly for the CISO (median hourly net imports of 7 GW over the 2.5-year study period) and BPAT (median hourly net exports of 8 GW) regions. The median hourly exchange between the BPAT and CISO regions was only 1.7 GW, so most of their exchanges are actually with other regions in the WECC. The results shown in \autoref{fig:3a} are consistent with what was observed for demand, generation and total interchange in \autoref{fig:2}. Large adjustments are needed for the AZPS-SRP link. Adjustments are more significant for the flows reported by SRP according to \autoref{fig:3a} which can be explained in two ways: the raw data for the SRP-AZPS link is larger than that reported for the AZPS-SRP link, so the weight for the SRP-AZPS is smaller, and the normalization used in \autoref{fig:3a} is the median total interchange for each region, which is twice as large in AZPS as in SRP.

\par Overall, the quality of data on generation by source is higher than the quality of exchange data, as reported by \autoref{fig:4}. One notable exception is data from the Western Area Power Administration, Lower Colorado Region (WALC). Adjustments for generation from coal reported in \autoref{fig:4} are especially large. The more exhaustive figures in the supplemental reports once again provide deeper insight: data on coal generation was missing for most of the time span under consideration. In this instance, the reconciled data provided by our algorithm for coal generation in WALC should be interpreted as a credible first guess for what coal generation might have been in WALC during this time period, that is physically consistent with other pieces of information available on WALC. While the resulting time series should probably not be taken at face value, it can be used immediately by data users while they wait for updates from the U.S. EIA.

\subsection{Stress-testing the data reconciliation framework}
\label{sec:3.2}
\begin{figure}[tbhp]
\centering
\includegraphics[width=.95\textwidth]{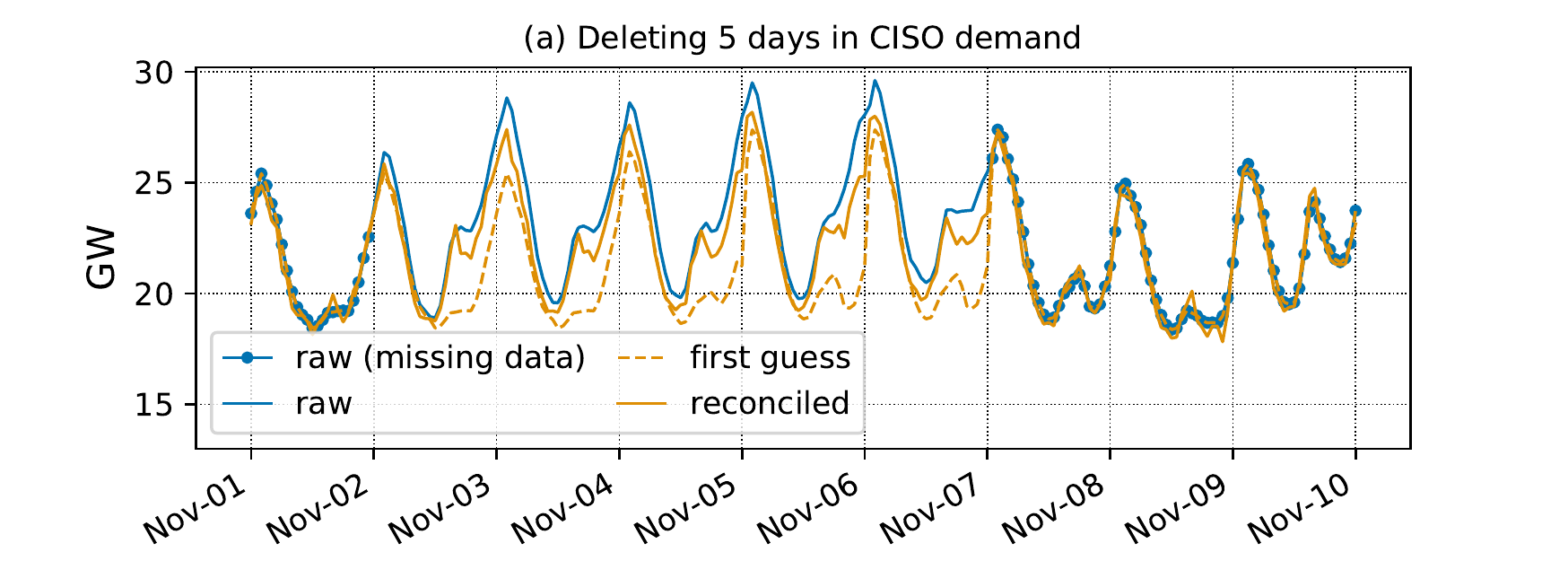}
\includegraphics[width=.95\textwidth]{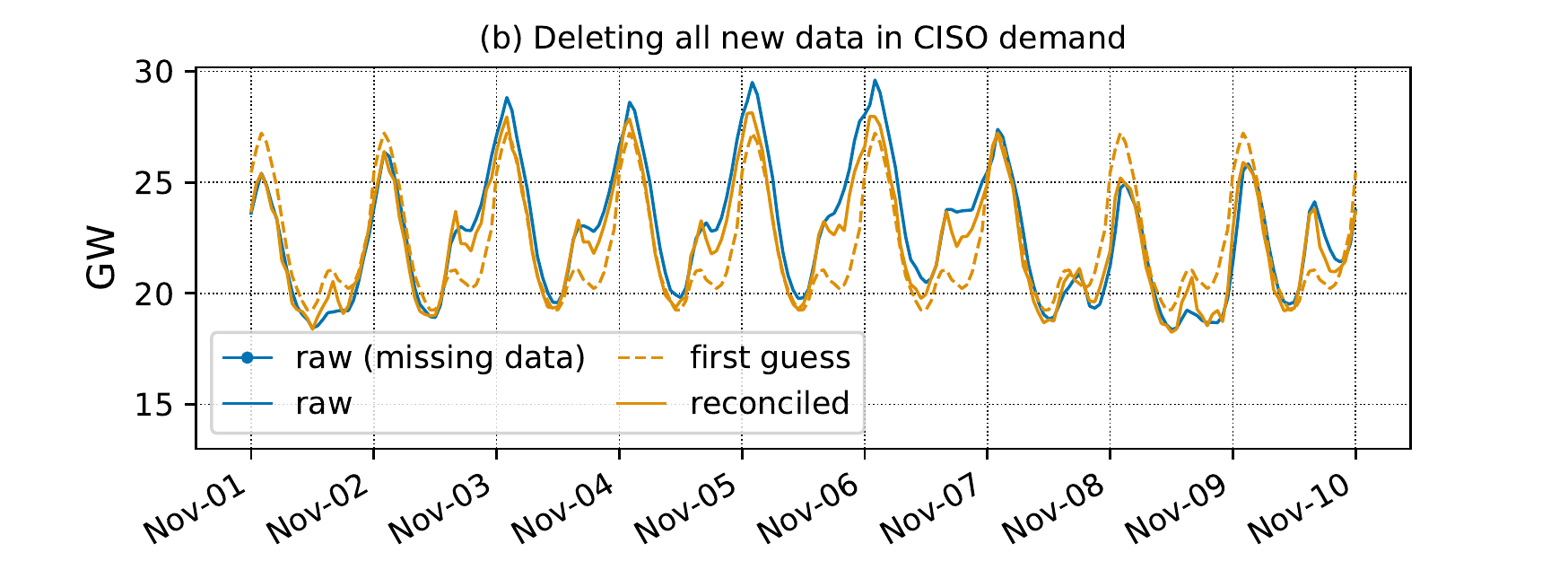}
\caption{\textcolor{mydiffcolor}{Robustness of the methodology to missing data. Missing values are artificially introduced for CISO demand from November 1\textsuperscript{st} to 10\textsuperscript{th} 2020. (a) Missing values are introduced for five days, data is available at the end of the horizon when providing a first guess in the data pre-processing step. (b) The entire time series to be reconciled is removed. Note that in this second case, historical data is still available during the data pre-processing step.}}
\label{fig:si_test1}
\end{figure}

\begin{figure}[tbhp]
\centering
\includegraphics[width=\textwidth]{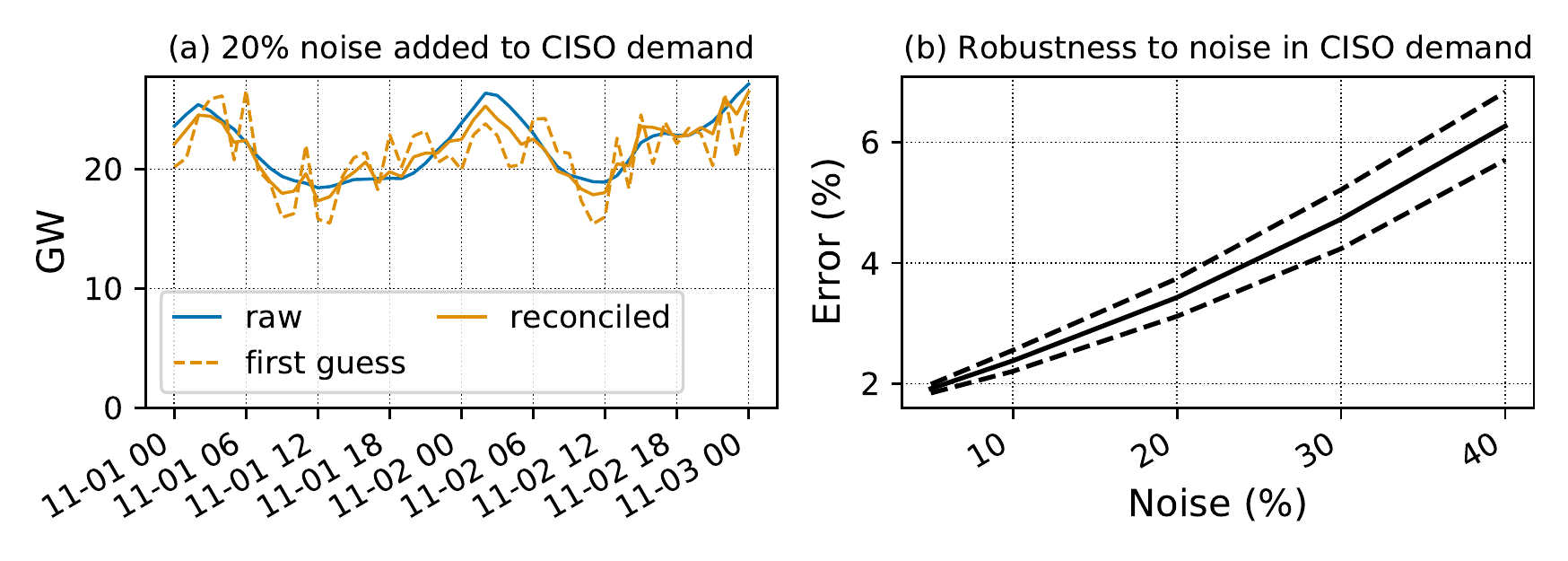}
\caption{\textcolor{mydiffcolor}{Robustness of the methodology to noisy data. Noisy values are artificially introduced in the time series for CISO demand from November 1\textsuperscript{st} to 3\textsuperscript{rd} 2020. The raw data set is multiplied by samples drawn from the uniform distribution $\mathcal{U}(1-l, 1+l)$, where the parameter $l$ controls the magnitude of the noise. The mean error is shown as a full black line, and a dashed black line denotes one standard deviation on either side of the mean. Statistics are computed using twenty full repetitions of each test.}}
\label{fig:si_test2}
\end{figure}

\textcolor{mydiffcolor}{
The ability of the methodology to deal with erroneous data is further assessed, first by introducing missing values in the data set, and then by adding noise to the data set. To construct the tests corresponding to the results shown in Figures~\ref{fig:si_test1} and~\ref{fig:si_test2}, demand data from the CISO was selected for a period where the raw data has good quality and can therefore be used as the ``source of truth''.}
\par \textcolor{mydiffcolor}{Figure~\ref{fig:si_test1} shows how the method performs with missing data. In the data pre-processing step, the procedure described in Section 2 is used to determine a first guess for replacing the missing values. The first guess is then adjusted by the data optimization step to meet the specified physical equations. This is the case for Figure~\ref{fig:si_test1} (a). When data is available only before the missing period, the last valid data for each hour of the day are propagated forward. This is the case for Figure~\ref{fig:si_test1} (b), where all new data for CISO demand were dropped, but historical data are available to provide a first guess. In both cases, the data optimization step is able to use information from the other data fields to refine the first guess and provides a suitable solution for the final reconciled data set as demonstrated by the good match between the reconciled and raw data.}
\par \textcolor{mydiffcolor}{In a second series of tests, noise is generated by repeatedly sampling from a uniform distribution over the interval $[1-l, 1+l]$, where $l$ is a parameter controlling the magnitude of the noise. Each demand value for CISO in the data set to be reconciled is multiplied by a different sample from this distribution. The resulting noisy data set does not have values that are unreasonable according to the data rejection heuristic used during the data pre-processing step, so the pre-processed data remain noisy, as shown in Figure~\ref{fig:si_test2} (a) for $l=.2$. Once again, the solution provided by the data optimization step greatly improves from the raw data. The errors in the reconciled data set are largest where the noise was largest. Figure~\ref{fig:si_test2} (b) reports on the results of repeated such tests for increasingly noisy data. The error reported in this figure corresponds to the mean absolute percentage error (MAPE) with respect to the raw data set, here used as ``source of truth''. Figure~\ref{fig:si_test2} (b) shows that the solution provided by the optimization program is robust to noise in the data, even under high levels of noise.}

\subsection{Continuously monitoring electricity and emissions}
\begin{figure}
\centering
\fbox{\includegraphics[width=\textwidth]{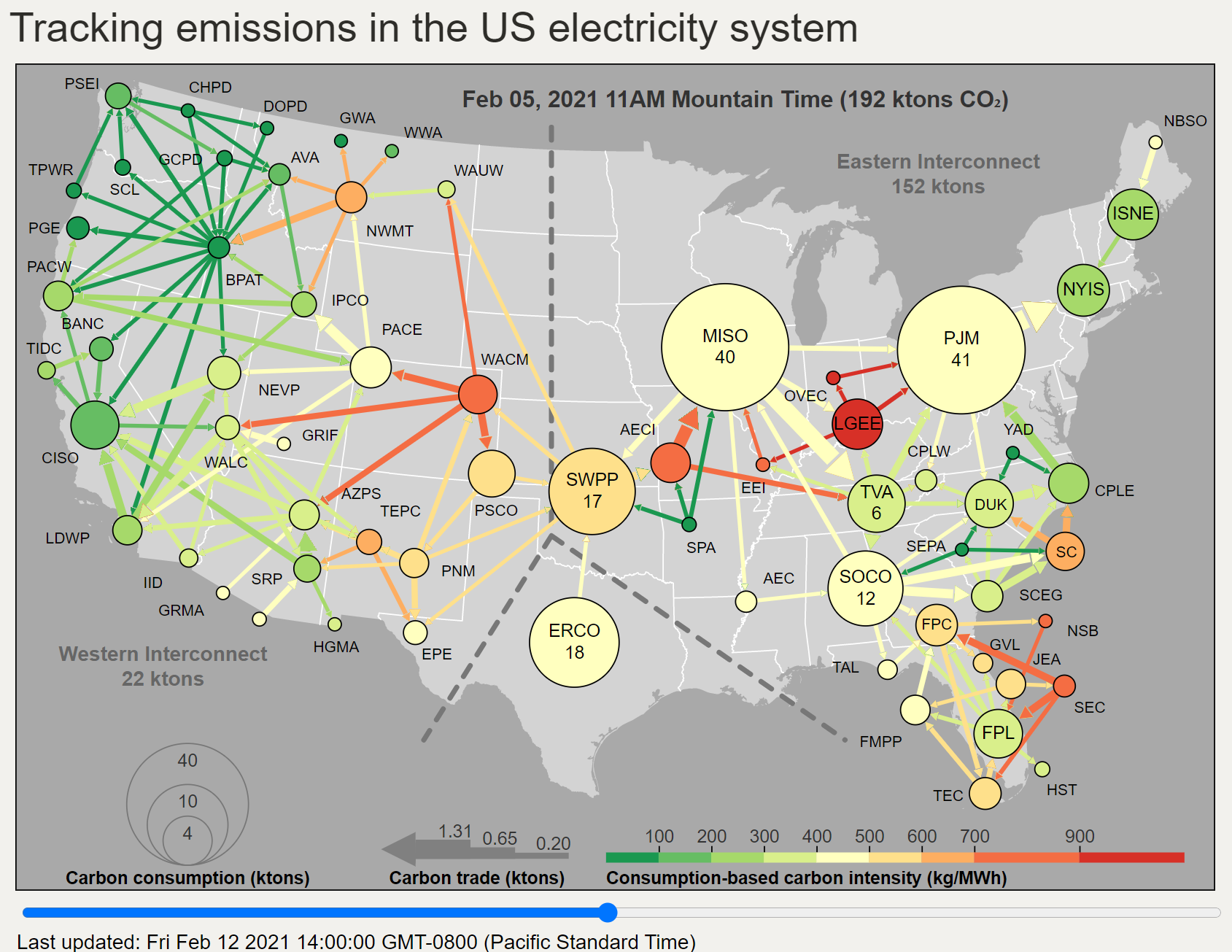}}
\caption{A real-time dashboard provides information on the carbon consumed and exchanged by the different regions in the US electricity system, here for February 5\textsuperscript{th}, 2021 at 11 AM (Mountain Time). New data is retrieved hourly from the EIA API, transformed using the process detailed in Section \ref{sec:2} and streamed to a publicly available location \cite{de2021data}. A reference for the acronyms in this map can be found online \cite{eia2021acronyms}.}
\label{fig:5}
\end{figure}

The automated data reconciliation framework developed in this work enables the development of near real-time tracking tools for electricity flows and the associated embodied emissions. One such tool, a real-time dashboard to monitor the carbon consumed and exchanged in the US electricity system, was developed in the context of this work and is illustrated in Figure~\ref{fig:5} \cite{de2021data}. Data transformed through the process described in Section~\ref{sec:2} is made publicly available at the same location for the benefit of researchers, policy makers and private sector actors so they can access up-to-date electricity and carbon data on the US electric grid. Reconciled data on electricity generation, consumption and exchanges is provided from January 2015 up to the present. Reconciled data on electricity generation by source and data on carbon emissions produced, consumed and exchanged is provided from July 2018. Figure~\ref{fig:5} illustrates how carbon consumption and the consumption-based carbon intensity of electricity vary in space. A slider in the web-based application can be used to explore how they vary in time.

\par The western interconnect consumed 13\% less carbon during daytime than during nighttime in the spring of 2019, while the eastern interconnect consumed 31\% more. The short version of this paper \cite{dechalendar2020recent} provides a discussion of trends seen in the data that are released. Nighttime wind power lowers the nighttime carbon intensity of Texas (ERCO) and the Southwest Power Pool (SWPP), while daytime solar lowers the daytime carbon intensity of the California regions. Abundant hydroelectric resources are behind the clean power that is produced and exported from the northwest of the US system. Exchanges play a crucial role in the western interconnection, while they play a much weaker role in the eastern US grid, where a few very large balancing authorities (PJM, MISO, SWPP, SOCO) account for a large fraction of total emissions.

\subsection{Impacts of the recent COVID-19 pandemic on the consumption of electricity and emissions in selected regions of the continental US}
As further illustration of the value of timely data on electricity and emissions, we provide a brief analysis of the impact of the COVID-19 pandemic on selected regions of the 2020 US electricity system. While several studies have already highlighted how the impact of the pandemic on electricity markets has varied across countries and regions \cite{buechler2020power, ruan2021quantitative}, especially in the February-May 2020 period, studies on emissions impacts in the electric sector have not been as highly spatially and temporally resolved \cite{le2020temporary}. The framework we provide is a key enabler for analyses of the emissions impacts of such disruptive events and for accelerating electric sector analyses in general.

\par
\begin{figure}[!tbh]
\centering
\includegraphics[width=\textwidth]{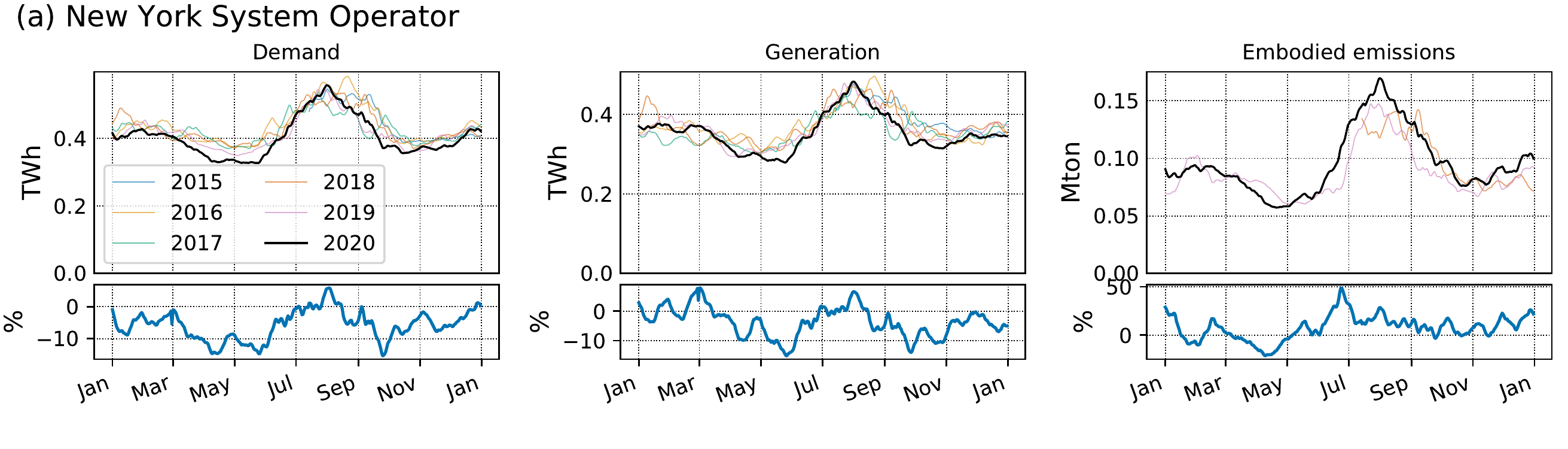}
\includegraphics[width=\textwidth]{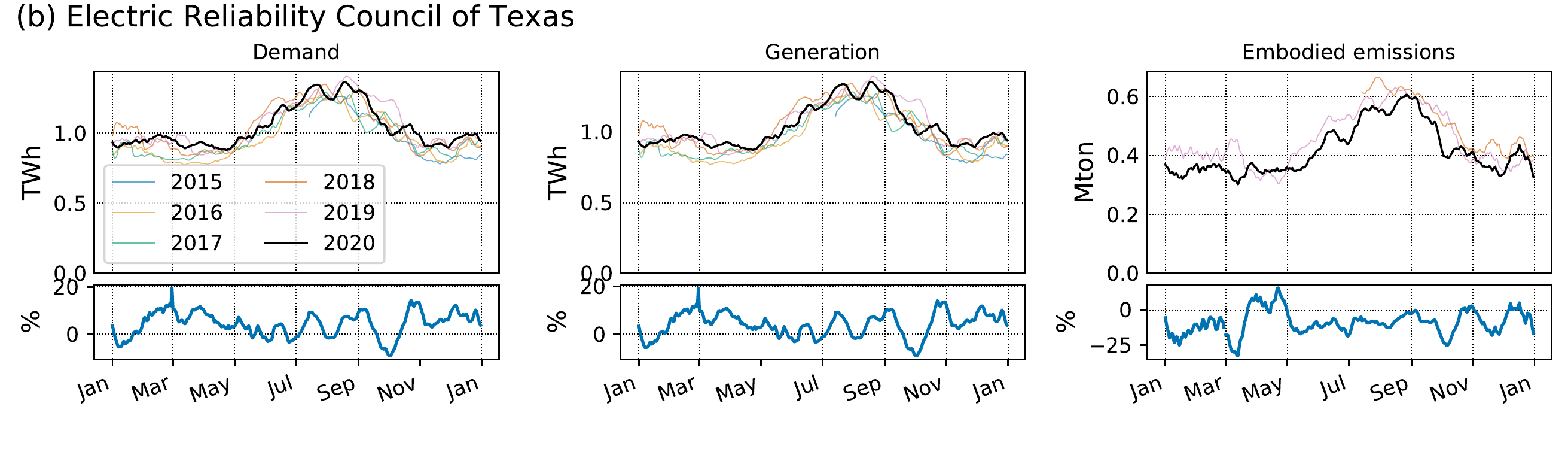}
\includegraphics[width=\textwidth]{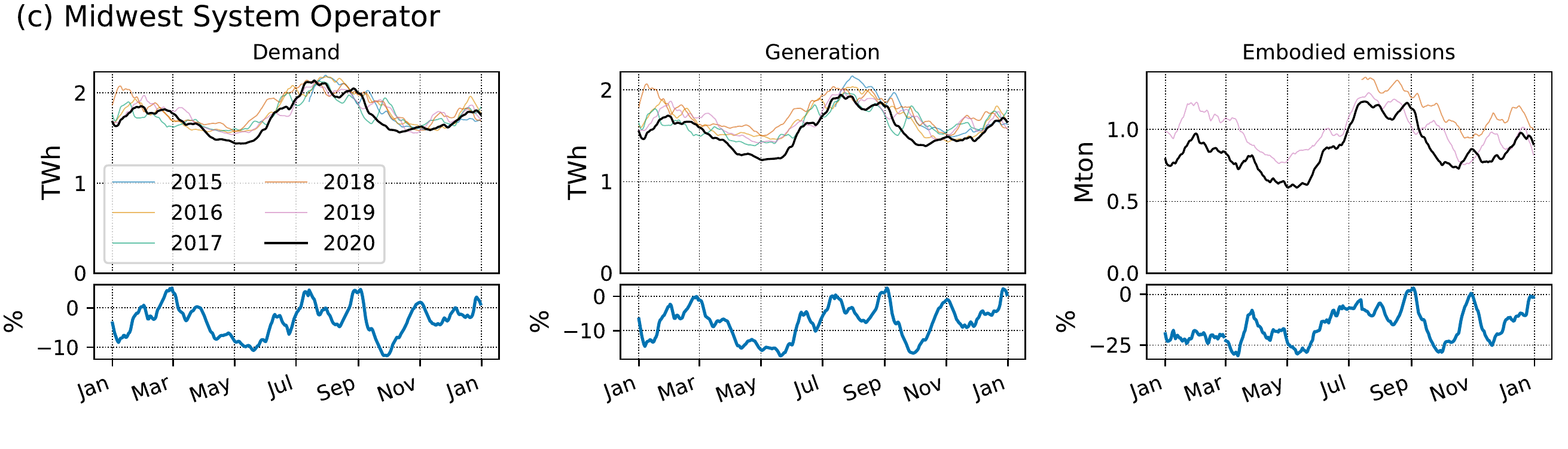}
\includegraphics[width=\textwidth]{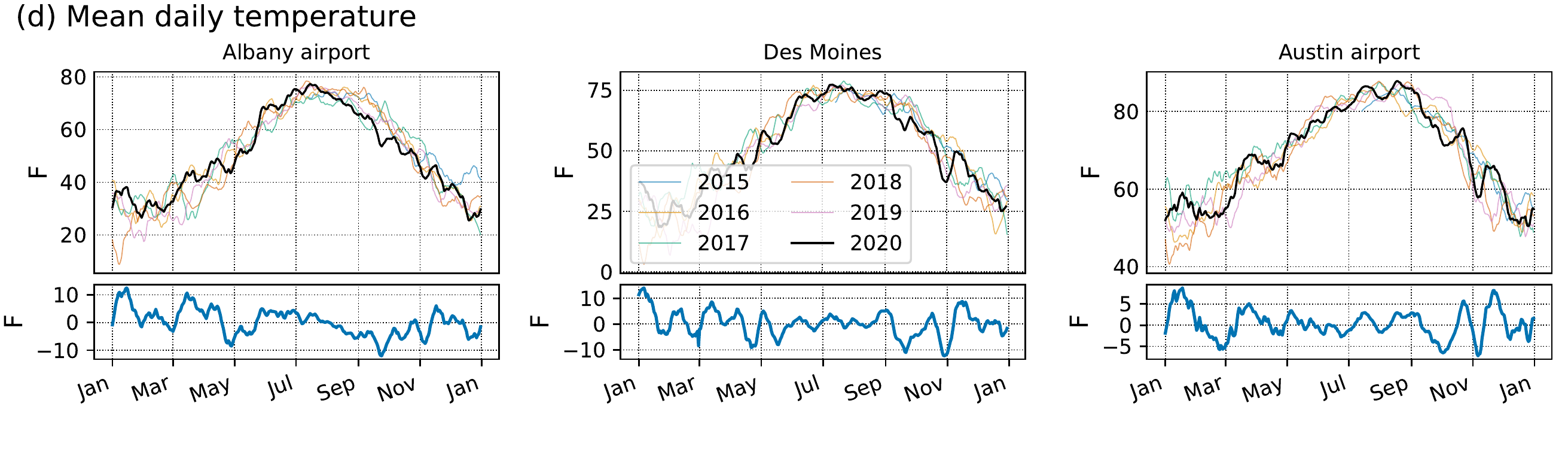}
\caption{Data shown is for (a) the New York System Operator (NYIS), (b) the Electric Reliability Council of Texas (ERCO) and (c) the Midwest System Operator (MISO). For each region, the top row shows the two-week rolling average for daily electricity demand and generation as well as embodied (consumed) emissions in different years, while the bottom row shows the percentage difference from the mean of 2016-2019 to 2020. Changes in emissions are measured using data for the second half of 2018 and 2019. (d) Two-week rolling average for mean daily temperature (top) and difference from the mean of 2018-2019 to 2020 (bottom) in Albany (New York), Des Moines (Iowa) and Austin (Texas).}
\label{fig:7}
\end{figure}

 \autoref{fig:7} presents data for the New York System Operator (NYIS), the Electric Reliability Council of Texas (ERCO) and the Midwest System Operator (MISO). For each region, the top row shows the two-week rolling average for daily electricity demand and generation as well as embodied (consumed) emissions in different years, while the bottom row shows the percentage difference from the mean of 2016-2019 to 2020. Changes in emissions are measured using data for the second half of 2018 and 2019. Temperature data from selected weather stations in each region are also shown in a similar format in \autoref{fig:7} (d).

\begin{figure}[!tbh]
\centering
\includegraphics[width=\textwidth]{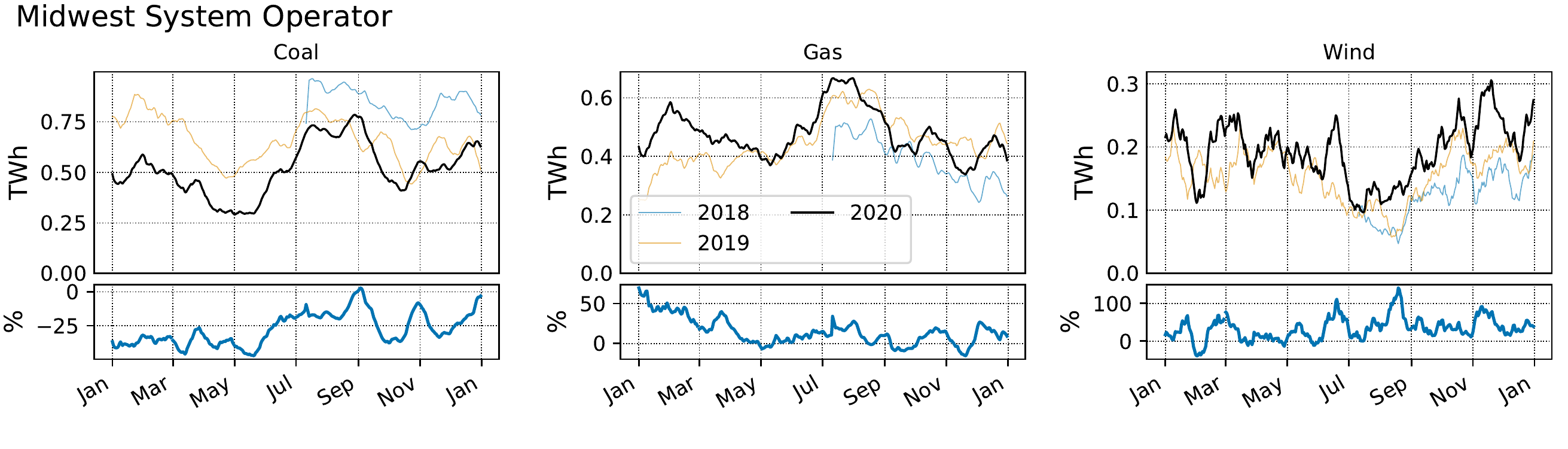}
\caption{Two-week rolling average for daily electricity production from coal, gas and wind (top row) and percentage difference from the mean of 2018-2019 to 2020 in the Midwest System Operator (MISO) region (bottom row).}
\label{fig:8}
\end{figure}

\par \textcolor{mydiffcolor}{Demand for electricity in NYIS demand began dropping quickly at the very beginning of March, consistent with the state of emergency declared in the state of New York on March 7\textsuperscript{th}. After several months of lower than usual demand, electricity demand rapidly picked up after New York City partially re-opened on June 8\textsuperscript{th}. Generation did not reduce as much, which can be explained by the fact that NYIS reduced their imports (from Canada, ISNE, PJM) instead of reducing generation in the balancing area. The changes in emissions observed during the first period of reduced demand are more than offset by increased emissions in the second part of the year. In the Texas grid (ERCO), the impacts of the pandemic on power demand and grid operations are less clear, with demand remaining well above historical levels throughout the year while emissions were far below historical levels.} 

\par In MISO, electricity demand started to drop in the second half of March and gradually went down by -10\% in June before recovering during the summer. Generation followed a similar pattern. Consumed emissions were down for the first quarter of 2021, before the pandemic started having any real impact, by close to -25\%, which can be explained by a large reduction in coal generation and an increase in wind as shown in \autoref{fig:8}. Coal production was down -25\% for most of the first half of the year, before recovering to normal levels. Wind generation was generally higher and represented 9\% of total 2020 generation in MISO. The lower coal output accounts for roughly half of the reduction in consumed emissions during the pandemic.

\par Temperature is a strong external driver for electricity demand, generation, and corresponding emissions. In all three regions shown in \autoref{fig:7}, for instance, a drop in electricity demand and generation (and emissions for ERCO and MISO) can be observed in early October 2020. As indicated by \autoref{fig:7} (d), lower than usual temperatures provide a good explanation for this drop that has a similar magnitude as the pandemic-related reduction earlier in the year, although the duration is much shorter. The impact of higher or lower than usual temperatures depends on the temperature range, as illustrated for instance by the temperature data for Austin, Texas: electric heating is impacted at lower temperatures, air conditioners are impacted at higher temperatures, and neither are on at milder temperatures. Correspondingly, higher or lower than usual temperatures do not have the same impact in February, October or November of 2020.

\section{Discussion}
The framework introduced in this paper offers a practical alternative that greatly enriches and complements the raw data set collected by entities such as the U.S. EIA. Researchers, policymakers and private sector actors with a interest in tracking electric sector emissions will benefit from this work. The methodology can be used both by those entities and data users (i) for anomaly detection and data validation and (ii) as an educated guess for data correction. This will be especially useful for practical applications where ensuring physical consistency in the data is important to the quality of the subsequent analyses.

\par The analysis of the performance of the method in section 3.1. shows how it can both provide a physically consistent first guess for data that provide information that conflicts with the information provided by other data fields, or for data that are missing. The method guarantees that the resulting data set will satisfy the equations enforced by optimization program (Eq.~\ref{eq:emissions_cleaning}). We note that these equations can be modified for applications that have different consistency requirements and that the method can find application in other fields. More generally, the method provides a good practical solution to the problem of tracking data inconsistencies in systems with sensors that provide redundant information.

\par \textcolor{mydiffcolor}{The method does not, however, guarantee that the resulting data set will be accurate, \textit{e.g.} it will not provide accurate information when two data fields provide erroneous data that are consistent with each other. Two underlying assumptions are that 1) at least one of the input data is correct and 2) that the errors in incorrect data remain relatively small. In the case where the second assumption does not hold, the method still provides value to identify inconsistent data fields, but the resulting data corrections will likely not provide reliable information.}

\par The analysis complements the data quality report that is released by the U.S. EIA \cite{eia2019issues}. Beyond the identification of discrepancies between information from different data fields, the framework can provide data on the magnitude of the discrepancies as well as suggest options for correcting the data. When inconsistencies are relatively minor, the framework produces accurate data that can be used directly. When discrepancies are large, the adjustments that result from the reconciliation program can be used to diagnose the source of the discrepancies. In the U.S. EIA data set used for the illustration in this paper, data on imports and exports were found to have the lowest quality, compared to data on production, total generation and generation by source.

\par An important feature of the data reconciliation framework developed in this work is its flexibility. By default, the algorithm we use for the data reconciliation step does not ``choose'' one data source over another when there is a conflict between two sources of information (e.g. if two regions have different reports for exchanges between them). Rather, the weights that are automatically chosen in the data expansion step and the quadratic penalty we impose on data adjustments guide the algorithm towards reasonable compromises between conflicting sources of information. Users that have access to reliable information on a subset of data fields can effectively guide the algorithm towards their preferred solution by manually adjusting the corresponding data fields before the data reconciliation step and then supplying weights that place a high penalty on deviating from the data they provided.

\par The U.S. electricity system is slowly decarbonizing. Although much of the progress of the recent decade is attributable to gas-fired generators replacing traditional coal-fired US power plants, renewables have also steadily been gaining ground. The monitoring tools developed in this work show how the carbon intensity of consumed electricity varies in time and in space in electric grids with significant penetrations of renewable generation. Continuously tracking embodied emissions flows will be critical to monitor decarbonization progress and direct climate policy to when, and where, it is most useful. As installed capacity for renewables grows, it is likely electricity exchanges over the transmission lines that connect different regions of the power grid will too. Tracking tools to monitor electricity and carbon flows will become increasingly useful.

\par Performing in-depth analyses and characterizations of the way the different electricity grids are adapting their operations to renewables will also be critical to push the decarbonization frontier. Such analyses will provide valuable insights to grids that are exploring decarbonization pathways or have not yet committed to a climate strategy, as well as to regions of the world where electricity grids have yet to be built.

\section{Conclusion}
The main contribution of this work is a physics-informed data reconciliation framework that provides a solution to the general problem of correcting data inconsistencies in systems with sensors that provide redundant information. In the context of electricity networks, the framework relies on solving an optimization program formed with the data adjustments needed for the corrected data set to satisfy energy conservation equations.
\par The performance of the framework is assessed by applying it to historical data from the continental United States electricity network; emissions for electricity consumption, production and exchanges are also computed. Two illustrative examples are used to show how the information can be used to provide valuable insights. The resulting data set on electricity and emissions is made publicly available and updated hourly. The dataset will provide valuable up-to-date carbon intensity information to private sector actors and policy makers. The method that was used to generate the dataset will benefit researchers in the field and in related fields that require access to such physically consistent data.

\section*{Acknowledgments}
Funding for this research was supported by a State Grid Graduate Student Fellowship through the Stanford Bits \& Watts initiative and by TotalEnergies.

\bibliography{mybibfile}

\end{document}